\documentclass[aps,prl,twocolumn,groupedaddress,amsmath,superscriptaddress,showpacs]{revtex4-1}

\usepackage{amsfonts}
\usepackage{amssymb}
\usepackage{amsmath}
\usepackage[dvips]{graphicx}
\usepackage{color}
\usepackage{bbm}
\usepackage{ulem}
\definecolor{nblue}{rgb}{0.3,0.3,1.0}
\definecolor{ngreen}{rgb}{0.2,0.7,0.2}
\definecolor{nred}{rgb}{0.9,0.1,0}
\definecolor{nblack}{rgb}{0,0,0}

\usepackage{amsmath}
\usepackage{graphicx}
\usepackage{amsfonts}
\usepackage{amssymb}
\usepackage{hyperref}
\usepackage{color}
\usepackage{mathrsfs}
\usepackage{isomath}
\usepackage{amsmath}
\usepackage{amsthm}
\usepackage{epstopdf}
\usepackage{txfonts}
\usepackage{cancel}

\newcommand{\ket}[1]{\left|#1 \right\rangle}
\newcommand{\bra}[1]{\left\langle#1 \right|}

\definecolor{maroon}{rgb}{0.7,0,0}

\begin{document}
	
	\title{Experimental Test of Contextuality based on State Discrimination with a Single Qubit}  
	
	\author{Qiuxin Zhang}
	\address{Department of Physics, Renmin University of China, Beijing 100872, China}
	\author{Chenhao Zhu}
	\address{Department of Physics, Renmin University of China, Beijing 100872, China}
	\author{Yuxin Wang}
	\address{Department of Physics, Renmin University of China, Beijing 100872, China}
	\author{Liangyu Ding}
	\address{Department of Physics, Renmin University of China, Beijing 100872, China}
	\author{Tingting Shi}
	\address{Department of Physics, Renmin University of China, Beijing 100872, China}
	\author{Xiang Zhang}
	\address{Department of Physics, Renmin University of China, Beijing 100872, China}
	\address{Beijing Academy of Quantum Information Sciences, Beijing 100193, China}
	\author{Shuaining Zhang}
	\email{zhangshuaining@ruc.edu.cn}
	\affiliation{Department of Physics, Renmin University of China, Beijing 100872, China}
	\affiliation{Beijing Academy of Quantum Information Sciences, Beijing 100193, China}
	\author{Wei Zhang}
	\email{wzhangl@ruc.edu.cn}
	\address{Department of Physics, Renmin University of China, Beijing 100872, China}
	\address{Beijing Academy of Quantum Information Sciences, Beijing 100193, China}
	\address{Beijing Key Laboratory of Opto-electronic Functional Materials and Micro-nano Devices, Renmin University of China, Beijing 100872, China}
	
	\begin{abstract}
		
		Exploring quantum phenomena beyond predictions of any classical model has fundamental importance to understand the boundary of classical and quantum descriptions of nature. As a typical property that a quantum system behaves distinctively from a classical counterpart, contextuality has been studied extensively and verified experimentally in systems composed of at least three levels (qutrit). Here we extend the scope of experimental test of contextuality to a minimal quantum system of only two states (qubit) by implementing the minimum error state discrimination on a single $^{171}$Yb$^+$ ion. We observe a substantial violation of a no-go inequality derived by assuming non-contextuality, and firmly conclude that the measured results of state discrimination cannot be reconciled with any non-contextual description. We also quantify the contextual advantage of state discrimination and the tolerance against quantum noises.
	\end{abstract}
	
	\maketitle
	
	\textit{Introduction.--} 
	In 1935, Einstein, Podolsky, and Rosen~\cite{EPR35} firstly posed the question of whether quantum observables have physical reality. It is then realized that such a question has significant physical indication rather than a conceptual or metaphysical notion. A variety of statements called {\it no-go theorems} have been formulated in mathematical terms to exclude from our natural world some different assumptions of reality. The pioneering works by Bell~\cite{Bell65,BrunnerRMP} and Kochen and Specker (KS)~\cite{Kochen,contextreview} provide two successful quantitative descriptions, where the violation of Bell inequality effectively precludes local causation, and the no-go theorem of KS theory proves that quantum mechanics will give conflicting predictions from any classical non-contextual model~\cite{Kochen,contextreview}. These statements are of great importance since they are both rigorous in mathematics and realistic in experiments~\cite{PRA4}, where the test of Bell inequality usually requires entangled states, and KS inequality can be violated for a more general state of a quantum system.
	
	The test of KS inequality has been demonstrated in various physical platforms composed of photons~\cite{photons1,photons2}, superconducting systems~\cite{NCSuperconduct},  nuclear spins~\cite{NVCenter}, ions~\cite{ions1,ions3,context1}, or neutrons~\cite{neutrons1}. In fact, in the conventional formalism of KS theory, a {\it qutrit} system including at least three levels is the simplest case to show quantum contextuality. Previous works are then demonstrated either for two-qubit systems or a qutrit with multiple observables~\cite{NVCenter,ions1,Mermin,Peres,Cabello,contextnature, ions3,NCSuperconduct,photons1, neutrons1, photons2, kcbs, Cabellokcbs, context1}. Recently, a research about contextual advantage for quantum state discrimination (SD) suggests that a test of contextuality may be extended to even simpler systems and with fewer observables~\cite{PRX}.
	
	In this work, we show the first experimental test of quantum contextuality in the simplest possible system of {\it a single qubit} with only {\it two observable measurements} based on the non-classicality of state discrimination. State discrimination focuses on how to effectively determine the state of a quantum system according to a set of measured probabilities with a known prior distribution. For a Hermitian system, one can prove that two non-orthogonal pure states cannot be entirely discriminated within either a classical or a quantum theory. However, a quantum description can significantly improves the probability of success, such that the research along this line is of great interest in both the study of fundamental quantum physics~\cite{PBR} and the search of potential applications in quantum information processing~\cite{QKD, QIP}. This advantage occurs exclusively in non-classical theory and has been studied in various areas such as quantum clone theorem~\cite{NJP1,NJP2,NJP6}, dimension witness~\cite{NJP18}, and mutual information~\cite{NJP19}. Using a trapped $^{171}$Yb$^+$ ion, we realize for the first time an experimental test of contextuality under the minimal error state discrimination (MESD) scenario with equal prior probability, where a no-go inequality that tolerates experimental error in non-contextual notation can be derived~\cite{PRX}. With high-fidelity single-qubit logic gates and highly distinguishable state-dependent fluorescence detection, a significant violation of the inequality is observed and the experimental results are well agreed with theoretical predictions of a quantum description. 
	
\begin{figure}[tbp]
		\centering
		\includegraphics[width=\linewidth]{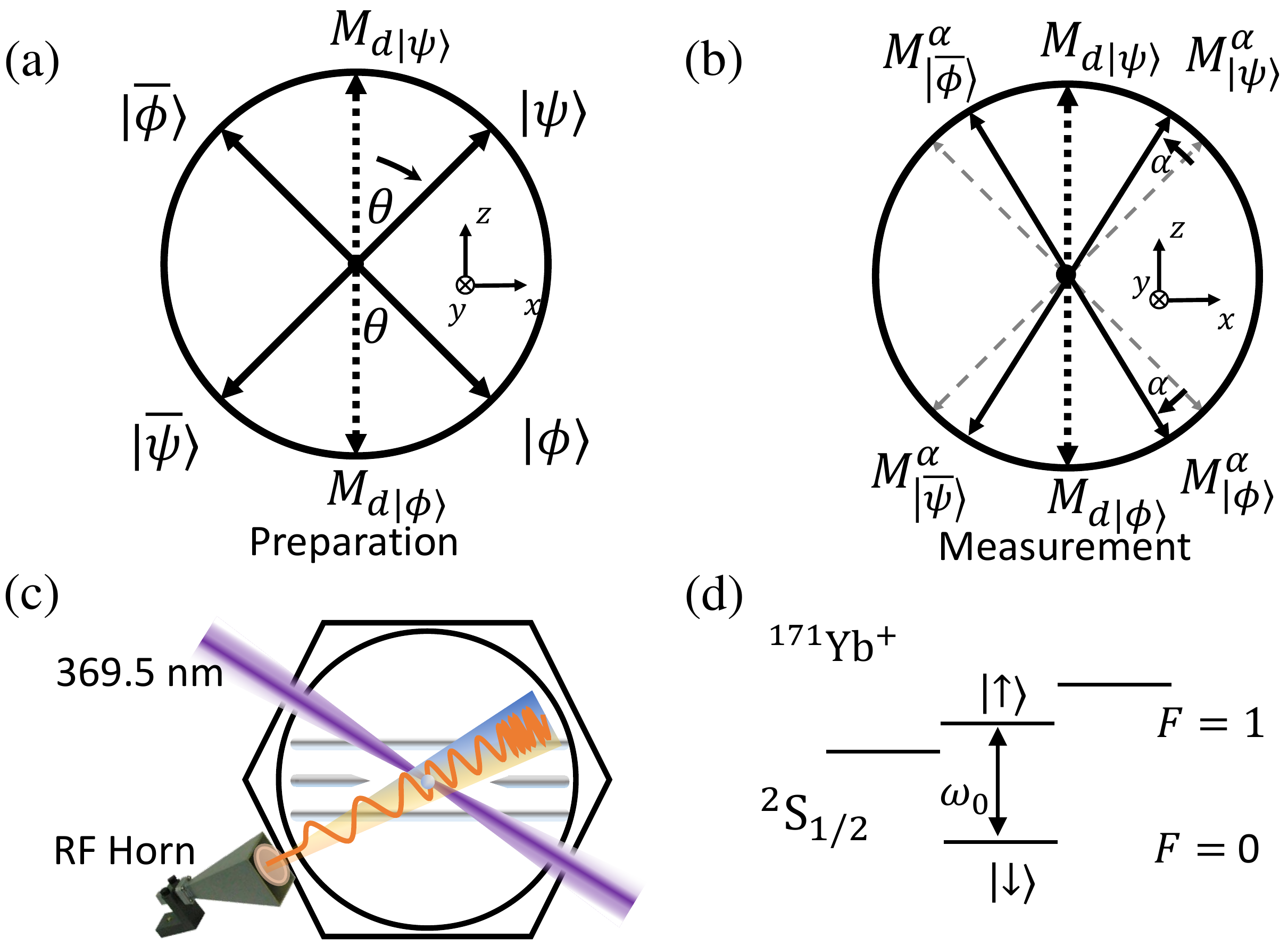}
		\caption{
			(a) The scheme of state preparations on the $x$-$z$ Bloch plane. The state $\ket{\psi(\phi)}$ (dark arrow) has an angle $\theta$ with measurement basis $M_{d\ket{\psi(\phi)}}$ (dashed line). We change the value of $\theta$ from 0 to $\pi/2$ to prepare different state ensembles, such that the two states change from orthogonal to parallel. The state $\ket{\bar{\psi}(\bar{\phi)}}$ is orthogonal to the state $\ket{\psi(\phi)}$.
			(b) The measurement scheme for states on the $x$-$z$ Bloch plane. The ideal measurement basis are shown with gray dashed lines. A measurement error $\epsilon$ is considered in the basis $M^\alpha_{\ket{\psi(\phi)}}$($\alpha \in [0,\pi/2]$), which has an angle deviation $\alpha$ from the ideal basis $M_{\ket{\psi(\phi)}}$. The measurement basis $M^\alpha_{\ket{\phi(\psi)}}$ and $M^\alpha_{{\ket{\bar{\phi}(\bar{\psi})}}}$ are orthogonal.
			(c) The trapped single $^{171}$Yb$^+$ ion system with a four-rod Paul trap. The purple beam is the 369.5nm laser for cooling, state initialization and state readout. The single-qubit gate is realized by a microwave horn antenna.
			(d) The energy level diagram of $^{171}$Yb$^+$.
		}
		\label{fig1}
	\end{figure}

	
	\textit{Preliminary.--} We consider the problem of distinguishing two pure states $\ket{\psi}$ and $\ket{\phi}$ with equal priori probability, as shown in Fig.~\ref{fig1}(a). The states $\ket{\bar{\psi}}$ and $\ket{\bar{\phi}}$ are orthogonal to $\ket{\psi}$ and $\ket{\phi}$, respectively, and are introduced as complement to construct operational equivalence \cite{PRA4} (See {\it Data analysis and results}) and derive the no-go inequality with the hypothesis of non-contextuality.  By defining the Helstrom measurement~\cite{Helstrom} basis set (the dashed line) as $\{M_{d\ket{\psi}}=1/2(I+\sigma_z), M_{d\ket{\phi}}=1/2(I-\sigma_z)\}$ with $M_{d{\ket{\psi}}} + M_{d\ket{\phi}}=I$, we can write the probability of a successful differentiation between $\ket{\psi}$ and $\ket{\phi}$ as $s=\frac{1}{2}({\rm Tr}[M_{d\ket{\psi}}\ket{\psi}\bra{\psi}]+{\rm Tr} [M_{d\ket{\phi}}\ket{\phi}\bra{\phi}])$. It is reasonable and inevitable for maximizing the successful probability to assume that the measurements have the symmetry property ${\rm Tr}[M_{d\ket{\psi}}\ket{\psi}\bra{\psi}]={\rm Tr} [M_{d\ket{\phi}}\ket{\phi}\bra{\phi}]$,  so that $s={\rm Tr}[M_{d\ket{\psi}}\ket{\psi}\bra{\psi}]={\rm Tr} [M_{d\ket{\phi}}\ket{\phi}\bra{\phi}]$. With that, the quantity $s$ is related to the overlap of $\ket{\phi}$ and $\ket{\psi}$ as
	\begin{eqnarray}
		s =\frac{1}{2}\left(1+\sqrt{1-|\langle\phi|\psi \rangle|^2}\right) =\frac{1}{2}\left(1+\sqrt{1-c}\right)
		\label{quantumS},
	\end{eqnarray}
	where the overlap $c \equiv |\langle\phi|\psi \rangle|^2$ is usually referred as confusability. It is obvious that when the direction of the measurement basis $M_{d\ket{\psi(\phi)}}$ and the direction of the vector $(\ket{\psi}-\ket{\phi})$ are parallel as shown in Fig.~\ref{fig1}(a), the error rate of the two states discrimination is the lowest, by which we call the minimum-error state discrimination (MESD).

	As proved in Ref.~\cite{PRX}, any ontological model of an operational theory with non-contextuality must lead to the following no-go theorem for MESD 
	\begin{eqnarray}
		s\leq 1-\frac{c}{2}.
		\label{noncS1}
	\end{eqnarray}
	This inequality would be relaxed owing to the error of quantum operations or measurement basis under realistic experimental conditions. To this end, one obtains 
	\begin{eqnarray}
		s\leq 1-\frac{c-\epsilon}{2},
		\label{noncS}
	\end{eqnarray}
	where $\epsilon$ is a quantitative indicator of measurement incorrectness, which is introduced just like the one appeared in the Clauser-Horne-Shimony-Holt (CHSH) inequality~\cite{CHSH} to describe the error of measurement in the Bell inequality. In our experiment, $\epsilon$ is related to the deviation of the actual measurement basis $\{M_{\ket{\psi}}^\alpha, M_{\ket{\phi}}^\alpha\}$ [solid lines in Fig.~\ref{fig1}(b)] from the ideal measurement axes (dashed lines) with an angle $\alpha \in [0,\pi/2]$. We further choose the directions of deviation for $M_{\ket{\psi}}^\alpha$ and $M_{\ket{\phi}}^\alpha$ are opposite to each other, i.e., they are tilted either away or close to the $x$-axis simultaneously.
	
	Taking the measurement error into account, the quantum bound of success probability of SD can be derived as~\cite{PRX}
	\begin{eqnarray}
		s=\frac{1}{2} \left[1+\sqrt{1-\epsilon+2 \sqrt{\epsilon(1-\epsilon) c(1-c)}+c(2 \epsilon-1)} \right].
		\label{quantumES}
	\end{eqnarray}
	Apparently, the expression of Eq.~(\ref{quantumES}) reduces to that of Eq.~(\ref{quantumS}) when $\epsilon=0$, and the no-go result of Eq.~(\ref{noncS1}) is the corresponding limiting case of Eq.~(\ref{noncS}). The success probability $s$ of discrimination in quantum description is higher than the upper bound of non-contextual theories. We use the same terminology as in Ref.~\cite{PRX} to label the measurements and outcomes in a natural way. For instance, the outcome of $M_{\ket{\phi}}^\alpha$ is more likely to occur than the outcome of $M_{\ket{\psi}}^\alpha$ if one begins with $P_{\ket{\phi}}$ rather than $P_{\ket{\bar{\psi}}}$ preparation, and so on. Thus, Eqs.~(\ref{noncS}) and (\ref{quantumES}) must satisfy the constraint
	\begin{eqnarray}\label{cons}
		\epsilon \leq c \leq 1-\epsilon.
	\end{eqnarray}
	We also calculate the theoretical predications of $c=\langle \psi| M^\alpha_{\ket{\phi}}|\psi\rangle$, $s=\langle \phi| M_{d\ket{\phi}}|\phi\rangle$, and $1-\epsilon=\langle \phi| M^\alpha_{\ket{\phi}}|\phi\rangle$ under a quantum description for different state preparations and measurements. The precise predictions are listed in Table~\ref{table1}. Note that two additional preparations along the $y$-axis are introduced as an extra dimension to correct the experimental preparation inaccuracy and supplement the measurement observable $\sigma_y$ to ensure the completeness of tomographic measurement.

	\begin{table}[tp]
		\centering
		\caption{Theoretical predictions for different state preparations and measurements. Each pair of preparation $P_{i}^\theta$ (with $i\in\{\ket{\phi},\ket{\bar{\phi}},\ket{\psi},\ket{\bar{\psi}\}}$) and measurements$M_{j}$ (with $M_j \in\{M_{\ket{\psi}}^\alpha, M_{\ket{\phi}}^\alpha, M_{d\ket{\phi}}\}$) yields a set of parameters $c$ = $\sin^2[({\alpha-2 \theta})/{2}]$, $\epsilon$ = $ \sin^2({\alpha}/{2})$ and $s$ = $\cos^2({\theta}/{2})$. To ensure that the preparations and measurements are tomographically complete, we add the preparations \{$P_y$,$P_{\bar{y}}$\} and measurement $M_y$ in the $y$-axis.}
		\label{table1}
		\small
		\resizebox{0.28\textwidth}{!}{%
			\begin{tabular}{ccccccc}			
				\hline\hline
				\textbf{} &
				\textbf{$P_{\ket{\psi}}^\theta$} &
				\textbf{$P_{\ket{\bar{\psi}}}^\theta$} &
				\textbf{$P_{\ket{\phi}}^\theta$} &
				\textbf{$P_{\ket{\bar{\phi}}}^\theta$} &
				\textbf{$P_y$} &
				\textbf{$P_{\bar{y}}$} \\ \hline
				\textbf{$M_{\ket{\psi}}^\alpha$} & $1-\epsilon$            & $\epsilon$            & $c$ & $1-c$ & 0.5 & 0.5 \\ 
				\textbf{$M_{\ket{\phi}}^\alpha$} & $c$ & $1-c$ & $1-\epsilon$            & $\epsilon$            & 0.5 & 0.5 \\ 
				\textbf{$M_{d\ket{\phi}}$} &
				$1-s$ &
				$s$ &
				$s$ &
				$1-s$ &
				0.5 &
				0.5 \\  
				\textbf{$M_y$}    & 0.5                          & 0.5                          & 0.5                          & 0.5                          & 1   & 0   \\ \hline\hline
			\end{tabular}%
		}
	\end{table}
	To test the no-go inequality Eq.~(\ref{noncS}), we need to prepare all states and measure all observables as given in Table~\ref{table1}, under the condition of operational equivalence. Then we can obtain the values of $\{ c, \epsilon, s \}$ from the relations given above and substitute into Eq.~(\ref{noncS}). If a violation of such an inequality is observed beyond experimental error, one can conclude that no non-contextual model can account for the experimental observation. Obviously this situation will be more difficult to occur with increasing noise. In this sense, the no-go inequality Eq.~(\ref{noncS}) provides a criterion for the boundary between quantum and classical regimes.

	\textit{Experimental setup.--} The experiment is performed on a single $^{171}{\rm Yb}^{+}$ ion trapped in a four-rod Paul trap, as shown in Fig.~\ref{fig1}(c). A pair of clock states $\ket{F=0,m_F=0}$ and $\ket{F=1,m_F=0}$ of the $^2$S$_{1/2}$ ground-state manifold are used to encode a qubit as \{$\ket{\downarrow}$, $\ket{\uparrow}$\} with transition frequency $\omega_0 = 2\pi \times 12.64281$ GHz [Fig.~\ref{fig1}(d)]. The qubit can be initialized by Doppler cooling and optical pumping, and detected by shining a 369.5 nm laser to excite the state $\ket{\uparrow}$ to the $^2$P$_{1/2}$ manifold and collecting the fluorescent photons with a photo-multiplier tube (PMT)~\cite{Olmschenk}. The error of detecting $\ket{\uparrow}$ for $\ket{\downarrow}$ is 1.71\% and the other one is 2.08\%. We calibrate the results to remove the detection errors as proposed in Ref.~\cite{Detection}.  
	
	The states $\ket{\psi}$ and $\ket{\phi}$ that we want to prepare lie in the $x$-$z$ plane, and can be denoted as
	\begin{eqnarray}
		\begin{aligned}
			\ket{\psi}&=\operatorname{\cos}\left({\theta}/{2}\right)\ket{\downarrow}+ \operatorname{\sin}\left({\theta}/{2}\right)\ket{\uparrow},\\
			\ket{\phi}&=\operatorname{\cos}\left[{(\pi-\theta)}/{2}\right]\ket{\downarrow}- \operatorname{\sin}\left[{(\pi-\theta)}/{2}\right]\ket{\uparrow}
		\end{aligned}
	\end{eqnarray}
	on the basis of $\{\ket{\downarrow}, \ket{\uparrow}\}$. When $\theta$ changes from 0 to $\pi/2$, the two states become closer to the $x$-axis on the Bloch sphere with increasing confusability $c$. To prepare such states, single-qubit quantum gates are implemented via microwave with resonant frequency. By controlling the duration and phase of the microwave, we can achieve a native single-qubit rotation
	\begin{eqnarray}
		R(\beta, \gamma)=\left(\begin{array}{cc}
			\cos (\beta / 2) & -i e^{-i \gamma} \sin (\beta / 2) \\
			-i e^{i \gamma} \sin (\beta / 2) & \cos (\beta / 2)
		\end{array}\right),
	\end{eqnarray}
	where $\beta$ and $\gamma$ are the rotation angle and phase, respectively. The  $\pi$ time of the Rabi oscillation is 7.51 ${\mu}s$, corresponding to a Rabi frequency of $2\pi\times 66.6$ kHz.
	
	The experimental process is as follows: After  $500\ \mu s$ Doppler cooling, the qubit is initialized to the $\ket{\downarrow}$ state by a $6\ \mu s$ optical pumping. Then, the set of states $\{P_{\ket{\psi}}^\theta, P_{\ket{\bar{\psi}}}^\theta,P_{\ket{\phi}}^\theta, P_{\ket{\bar{\phi}}}^\theta\}$  are prepared through single-qubit rotations $\{R_y(\theta),R_y(\pi+\theta), R_y(\pi-\theta),R_y^\dagger(\theta)\}$, where $R_y(\theta)=R(\theta,\pi/2)$ and $R_y^\dagger(\theta)=R(\theta,3\pi/2)$. The observable set is chosen as $\{M_{\ket{\psi}}^\alpha, M_{\ket{\phi}}^\alpha,M_{d\ket\phi}\}$, which are realized by adding single-qubit operations $\{R_y^\dagger(\theta-\alpha),R_y^\dagger(\pi-\theta+\alpha), R_y^\dagger(\pi)\}$ to rotate the measurement basis from the original $z$-basis to the desired one. By scanning the parameter region $\{\theta, \alpha\}$, we can obtain the dataset  $\{c, \epsilon, s\}$. 
	
	The parameter $\epsilon$ is a function of the variable $\alpha$ as $\epsilon$ = $ \sin^2({\alpha}/{2})$. Similarly, we have $c$ = $\sin^2[({\alpha-2 \theta})/{2}]$. When $\epsilon$ has a certain value, $c$ must meet the inequality Eq.~(\ref{cons}), which can be reduced to 
	\begin{eqnarray}
		\sin^2\left({\alpha}/{2}\right) \leq \sin^2\left[({\alpha-2 \theta})/{2}\right] \leq \cos^2\left({\alpha}/{2}\right).
	\end{eqnarray}
	Note that $\sin^2[({\alpha-2 \theta})/{2}] \leq \cos^2({\alpha}/{2})$ is always true with $\theta,\alpha \in[0,\pi/2] $. Therefore, when $\alpha$ is fixed, the range of $\theta$ can be determined by solving $\sin^2({\alpha}/{2}) \leq \sin^2[({\alpha-2 \theta})/{2}]$ with $\theta,\alpha \in[0,\pi/2] $. In our experiment, the parameters $\theta$ and $\alpha$ are scanned with an interval of 0.1, resulting in 136 sets of data $\{c, \epsilon, s\}$. We repeat the experiment for 1000 trials for each choice of the data set.
	
	To correct the state preparation inaccuracy on the Bloch sphere and ensure the measurements are tomographically complete, we also supplement extra state preparations and observable measurements through the similar quantum operations. Details of this procedure will be given in the subsequent discussion and Supplemental Materials~\cite{supp}. As a result, the total state preparations are  $\{P_{\ket{\psi}}^\theta, P_{\ket{\bar{\psi}}}^\theta,P_{\ket{\phi}}^\theta, P_{\ket{\bar{\phi}}}^\theta, P_y, P_{\bar{y}}\}$, and observable measurements are $\{M_{\ket{\psi}}^\alpha, M_{\ket{\phi}}^\alpha, M_{d\ket{{\phi}}}, M_y \}$, as shown in Table \ref{table1}. 
	We emphasize that although more than two state preparations and observable measurements are carried in the experiment, only two observables with binary outcome are obtained from the set $\{M_{\ket{\psi}}^\alpha, M_{\ket{\phi}}^\alpha, M_{d\ket{{\phi}}} \}$, and are subsequently used to test the no-go inequality and show the boundary between quantum and classical regimes.

	\begin{figure}[tbp]
		\centering
		\includegraphics[width=\linewidth]{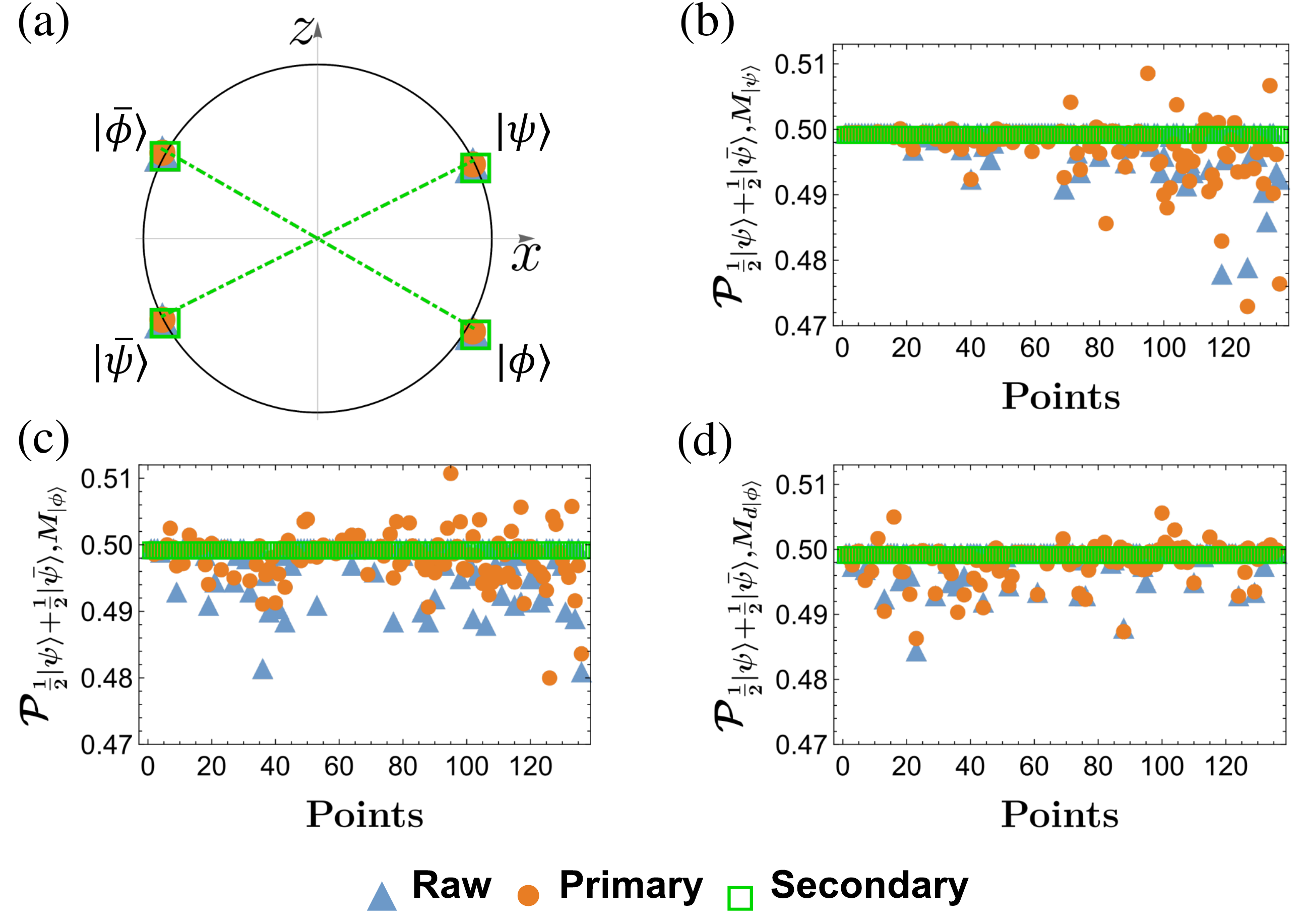}
		\caption{The results of data analysis for operational equivalence. 
			(a) The states $\{\ket{\phi}, \ket{\bar{\phi}}, \ket{\psi}, \ket{\bar{\psi}}\}$ prepared in experiments are plotted on the $x$-$z$ plane of the Bloch sphere with  $\theta$ = 1.1 rad and $\alpha$ = 0 rad, corresponding to the point with index 12 in panels (b), (c) and (d).
			Raw preparations (blue triangles) are given from the experimental data. Primary preparations (orange dots) are the results of fitting with the GPT theory. Secondary preparations (green rectangles) are chosen to satisfy the assumption of operational equivalence. 
			(b), (c), and (d) show that the probabilities $\mathcal{P}_{\frac{1}{2} \ket{\psi}+\frac{1}{2} \ket{\bar{\psi}}, M_{i}}$ reach statistical operational equivalence, i.e. $\mathcal{P}_{\frac{1}{2} \ket{\psi}+\frac{1}{2} \ket{\bar{\psi}}, M_{i}}=\frac{1}{2}$ after data processing. The results of $\mathcal{P}_{\frac{1}{2} \ket{\phi}+\frac{1}{2} \bar{\ket{\phi}}, M_{i}}$ are given in the Supplemental Materials.
		}
		\label{fig2}
	\end{figure}
	
	\textit{Data analysis and results.}--  The test of no-go inequality derived from the non-contextual model requires the data meet operational equivalence in a general framework regardless of quantum or classic. For this purpose, we need to process the raw experimental data in two stages~\cite{NC,Mikeprxq} (See details in the Supplemental Materials). The first phase fits the data of raw preparations using generalized probability theory (GPT) to ensure the states are not exclusively dependent on quantum theory under the condition of complete  tomographic measurements. The results of this stage, refereed as the primary preparations, are then supplied to the second phase which treats the data to perfectly comply to operational equivalence, i.e.
	\begin{eqnarray}
		\mathcal{P}_{\frac{1}{2} \phi+\frac{1}{2} \bar{\phi}, M_{i}}=\mathcal{P}_{\frac{1}{2} \ket{\psi}+\frac{1}{2} \ket{\bar{\psi}}, M_{i}}=\frac{1}{2}.
	\end{eqnarray}
	Here, $\mathcal{P}_{\frac{1}{2} \ket{\phi}+\frac{1}{2} \ket{\bar{\phi}}, M_{i}}$ denotes a process associated with a particular $M_i \in \{M_{\ket{\phi}}, M_{\ket{\psi}}, M_{d\ket{\phi}} \}$, which is randomly selected from the average random mid-sample of the $P_{\ket{\phi}}$ and $P_{\ket{\bar{\phi}}}$ preparation process in theory. The case is analogous for $\mathcal{P}_{\frac{1}{2} \psi+\frac{1}{2} \bar{\psi}, M_{i}}$. In experiment, $\mathcal{P}_{\frac{1}{2} \ket{\phi}+\frac{1}{2} \ket{\bar{\phi}}, M_{i}}$ is realized by a linear combination of the statistics of $\ket{\phi}$ and $\ket{\bar{\phi}}$ under measurement $M_{i}$, i.e.,  
	\begin{eqnarray}
		\mathcal{P}_{\frac{1}{2} \ket{\phi}+\frac{1}{2} \ket{\bar{\phi}}, M_{i}}=\frac{1}{2} \left[ P\left(1 \mid P_{\ket{\phi}}, M_{i}\right)+P\left(1 \mid P_{\ket{\bar{\phi}}}, M_{i}\right)\right].
	\end{eqnarray}
	Here, $P(1|P_{\phi(\bar{\phi})} ,M_i)$ denotes the frequency when the measurement axis is $ M_i$ and the outcome is 1. 
	Using maximum likelihood estimation method, we obtain the secondary state preparations by minimizing the target function as the summation of distances from their corresponding primaries.

	The two stages of data processing alter the position of states in the $x$-$z$ plane as illustrated in Fig.~\ref{fig2}(a). The data processing outcomes are plotted in Figs.~\ref{fig2}(b)-\ref{fig2}(d). Notice that the raw and primary data of state preparations are slightly off 1/2 owing to experimental errors in operation and measurement. The primary preparations fitted from the GPT theory have a slightly larger variance, which can be hardly observed in Figs.~\ref{fig2}(b)-\ref{fig2}(d). The results of secondary preparations show that all the 136 points fulfill the operational equivalence criterion with an exactly equal distribution 1/2. The similarity between states before and after data processing is evaluated as the average of mixture weights, which is determined as $97.13\% \pm 2.3\%$ in our experiment. Further details and results can be found in the Supplemental Materials.
	
	\begin{figure}[tbp]
		\centering
		\includegraphics[width=\linewidth]{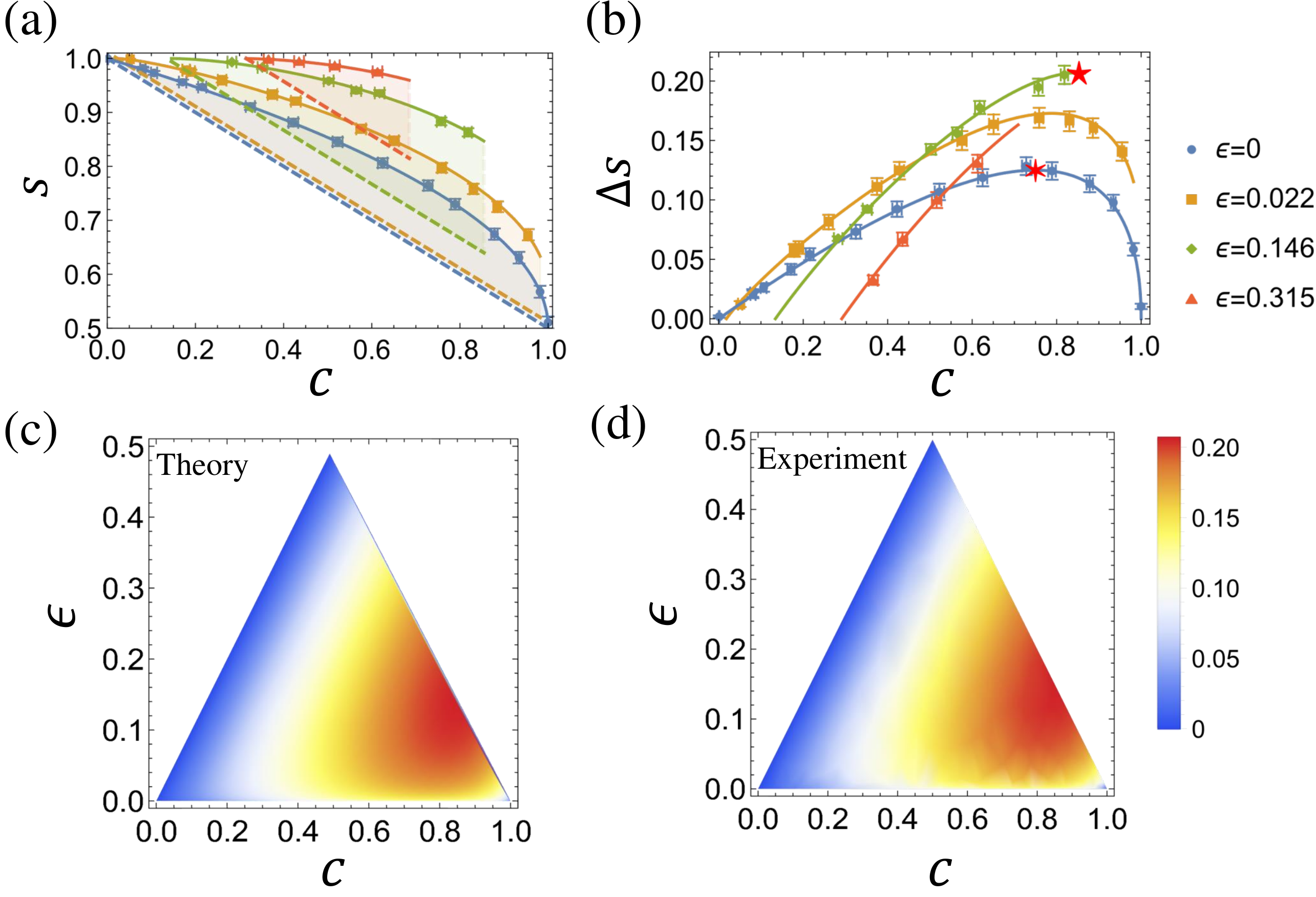}
		\caption{The results of contextual advantage for state discrimination.
			(a) The results of success probability $s$ versus confusability $c$ for $\epsilon$= $\{0, 0.022, 0.146, 0.315\}$. The points of dots, squares, diamonds and triangles are the experiment data. The dashed lines represent the non-contextual upper bound $s_{\rm non-con}$ given by Eq.~(\ref{noncS}), and the solid lines are the predictions by a quantum theory $s_{\rm theory}$. The shaded areas mark the contextual advantage $\Delta s_{\rm theory} = s_{\rm theory}-s_{\rm non-con}$ for state discrimination.
			(b) The results of $\Delta s$ versus $c$ with the same $\epsilon$'s and line shapes as in (a). The red pentagram and hexagonal star are the highest points with and without $\epsilon$, locating at ($c,\epsilon$)=(0.853, 0.146) and ($c$,$\epsilon$)=(0.75,0), respectively. (c) and (d) The false-color plots of $\Delta s$ by changing $c$ and $\epsilon$.}		
		\label{fig3}
	\end{figure}	
	
	We use the secondary preparation data which satisfies operational equivalence to calculate the results of $\{c,\epsilon,s\}$ and show the violation of the no-go inequality Eq.~(\ref{noncS}). In Fig.~\ref{fig3}(a), we plot the probability of successful state discrimination $s$ by varying the confusability $c$ with fixed measurement error $\epsilon= \{\sin (0/2)^2, \sin (0.3/2)^2, \sin (0.785/2)^2,\sin (1.2/2)^2\} = \{0, 0.022, 0.146, 0.315\}$. The measured $s_{\rm exp}$ is clearly above the upper bound $s_{\rm non-con} = 1-(c-\epsilon)/2$ for any non-contextual model (dashed lines), and agree well with theoretical predictions $s_{\rm theory}$ of Eq.~(\ref{quantumES}) given by quantum theory with contextuality (solid lines). The shaded region between the dashed and solid lines thus denotes the advantage of a contextual theory in SD. We also notice that in the limiting case of $c = \epsilon$, the success probability of SD can reach $s=1$ even for a pair non-orthogonal state preparations with $c >0$. 
	This is because in the presence of a finite $\epsilon$, a small confusability between two preparations with $c \le \epsilon$ can always be attributed to the measurement error, such that the two states are perfectly distinguished within an ontological description of the system, regardless of contextuality or non-contextuality.

	To show the contextual advantage of SD more clearly, we plot in Fig.~\ref{fig3}(b) the exceeding of $s$ over the non-contextual bound $\Delta s_{\rm theory/exp} = s_{\rm theory/exp}-s_{\rm non-con}$ for each choice of $\epsilon$. We emphasize that the highest value of $\Delta s_{\rm theory}=s_{\rm theory}-s_{\rm non-con}$ is 0.207 for the parameters ($c,\epsilon$)=(0.853, 0.146), which clearly exceeds the maximal value of 0.125 for the case with no measurement error achieved at ($c$,$\epsilon$)=(0.75,0). These two points are marked in Fig.~\ref{fig3}(b) with a red pentagram and a hexagonal star, respectively. It is obvious that the addition of a finite $\epsilon$ significantly improves the contextual advantage of SD over any non-contextual model. In experiment, the location of the pentagram and hexagonal star are determined at $\{c,\epsilon,s\}=\{0.750 \pm 0.010, 0.0000 \pm 0.0006, 0.125 \pm 0.007\}$ and $\{c,\epsilon,s\}=\{0.853\pm 0.009, 0.146\pm 0.011, 0.206\pm 0.006\}$, respectively. The quantity $\Delta s$ also manifests the tolerance against noise in SD. For instance, one can prove that in a contextual theory $s$ decreases with increasing depolarization error of state preparation and measurement~\cite{PRX}, which eventually compromises the violation of inequality Eq.~\ref{noncS}. Thus, a higher value of $\Delta s$ achieved at a finite $\epsilon$ indicates that a quantum SD is more robust against depolarization noises.
	
	The results of $\Delta s$ for all the 136 points obtained by theory and experiment are plotted in false color in Figs.~\ref{fig3}(c) and \ref{fig3}(d), respectively. To quantify the resemblance between theory and experiment, we define the average fidelity as
	\begin{eqnarray}
		f = 1-\frac{1}{N}\sum_{i=1}^{N=136} \frac{ \left| \Delta s_{{i, \rm exp}}-\Delta s_{i,{\rm theory}} \right| }{\Delta s_{i, {\rm theory}}},
	\end{eqnarray}
	and obtains $f = 98.82\%\pm 0.12 \%$. Since the gate infidelity induced by the fluctuation of microwave parameters is very low, which is $\sim (2.0\pm 0.3)\times$10$^{-4}$ as characterized by randomized benchmarking method~\cite{RB}, the main error source is the sampling error with an average of $1.29\%\pm 0.45\% $. (See details in the Supplemental Materials).

	\textit{Conclusion.}-- We demonstrate an experimental no-go test of contextuality derived within the frame work of minimum error state discrimination in a single qubit system with only two observable measurements. With high-fidelities of state initialization, manipulation and detection, we observe a probability of successful differentiation between two non-orthogonal state preparations which significantly exceeds the upper bound for any non-contextual theories. Our results represent the first experimental ruling out of non-contextual descriptions in a simple qubit system with only two observables. We also introduce the measurement error to quantify the contextual advantage and the tolerance over noise of quantum state discrimination. Our work not only deepen the understanding of the quantum nature of contextuality, but also motivate future research on noise effect from quantitative aspects.

	\begin{acknowledgments}
		We are thankful for support from the Beijing Natural Science Foundation (Grant No. Z180013), the National Natural Science Foundation of China (Grants No. 12074427, and No. 12074428), and the National Key R\&D Program of China (Grant No. 2018YFA0306501). S. Zhang also thanks the China Postdoctoral Science Foundation (Grant No. BX20200379 and No. 2021M693478) for support.
	\end{acknowledgments}



\begin{thebibliography}{99}
		
		\bibitem{EPR35} A. Einstein, B. Podolsky, and N. Rosen, Can quantum-mechanical description of physical reality be considered complete? Phys. Rev. \textbf{47}, 777 (1935).
		
		\bibitem{Bell65} J. S. Bell,  On the Einstein Podolsky Rosen paradox, Physics \textbf{1}, 195 (1964).
		
		\bibitem{BrunnerRMP} N. Brunner, D. Cavalcanti, S. Pironio, V. Scarani, and S. Wehner,  Bell nonlocality,  Rev. Mod. Phys. \textbf{86}, 419 (2014).
		
		
		\bibitem{Kochen} S. Kochen and E. P. Specker, The Problem of Hidden Variables in Quantum Mechanics, J. Math. Mech. \textbf{17}, 59 (1967).
		
		\bibitem{contextreview}C. Budroni, A. Cabello, O. G\"{u}uhne, M. Kleinmann, and J.-\AA{}. Larsson, Quantum contextuality,  arXiv:2102.13036.
		
		\bibitem{PRA4} W. Spekkens,  Contextuality for Preparations, Transformations, and Unsharp Measurements,  Phys. Rev. A \textbf{71}, 052108 (2005).
		
		\bibitem{photons1} A. Zhang, H. Xu, J. Xie, H. Zhang, B. J. Smith, M. S. Kim, and L. Zhang, Experimental test of contextuality in quantum and classical systems, Phys. Rev. Lett. \textbf{122}, 080401 (2019).
		
		\bibitem{photons2} D. Qu, K. Wang, L. Xiao, X. Zhan, and P. Xue,  State-independent test of quantum contextuality with either single photons or coherent light, NPJ Quantum Inf. \textbf{7}, 1 (2021).
		
		\bibitem{NCSuperconduct}  M. Jerger,  Y. Reshitnyk, M. Oppliger, A. Potočnik, M. Mondal, A. Wallraff, K. Goodenough, S. Wehner, K. Juliusson, N. K. Langford, and A. Fedorov, Contextuality without nonlocality in a superconducting quantum system, Nat. Commun.  \textbf{7}, 1 (2016).
		
		\bibitem{NVCenter}  S. B. van Dam, J. Cramer, T. H. Taminiau, and R. Hanso, Multipartite entanglement generation and contextuality tests using nondestructive three-qubit parity measurements, Phys. Rev. Lett. \textbf{123}, 050401 (2019).
		
		\bibitem{ions1} F. M. Leupold, M. Malinowski, C. Zhang, V. Negnevitsky, A. Cabello, J. Alonso, and J. P. Home, Sustained state-independent quantum contextual correlations from a single ion, Phys. Rev. Lett. \textbf{120}, 180401 (2018).
		
		\bibitem{ions3} P. F. Wang, J. H. Zhang, C. Y. Luan, Mark Um, Y. Wang, M. Qiao,  T. Xie, J. N. Zhang, A. Cabello, and K. Kim, Significant loophole-free test of Kochen-Specker contextuality using two species of atomic ions, Sci. Adv. \textbf{8}, eabk1660 (2022).
		
		\bibitem{context1} X. Zhang, M. Um, J. Zhang, S. An, Y. Wang, D.-l. Deng, C. Shen, L.-M. Duan, and K. Kim,  State-independent experimental test of quantum contextuality with a single trapped ion,  Phys. Rev. Lett.  \textbf{110}, 070401 (2013).
		
		\bibitem{neutrons1} Y. Hasegawa, R. Loidl, G. Badurek, M. Baron, and H. Rauch, Violation of a Bell-like inequality in single-neutron interferometry, Nature \textbf{425}, 45 (2003).
		
		\bibitem{Mermin} N. D. Mermin, Extreme quantum entanglement in a superposition of macroscopically distinct states, Phys. Rev. Lett. \textbf{65}, 1838 (1990).
		
		\bibitem{Peres} A. Peres, Two simple proofs of the Kochen-Specker theorem, J. Phys. A \textbf{24}, L175(1991).
		
		\bibitem{Cabello} A. Cabello, Experimentally Testable State-Independent Quantum Contextuality, Phys. Rev. Lett. \textbf{101}, 210401 (2008).
		
		\bibitem{contextnature}G. Kirchmair, F. Z\"{a}hringer, R. Gerritsma, M. Kleinmann, O. G\"{u}hne, A. Cabello, R. Blatt, and C. F. Roos,  State-independent experimental test of quantum contextuality,  Nature \textbf{460}, 494 (2009).
		
		\bibitem{kcbs} A. A. Klyachko, M. A. Can, S. Binicio\v{g}lu, and A. S. Shumovsky, Simple test for hidden variables in spin-1 systems,  Phys. Rev. Lett. \textbf{101}, 020403 (2008).
		
		\bibitem{Cabellokcbs}M. Kleinmann, C. Budroni, J-\AA{}. Larsson, O. G\"{u}hne, and A. Cabello, Optimal inequalities for state-independent contextuality, Phys. Rev. Lett. \textbf{109}, 250402 (2012).
		
		\bibitem{PRX} D. Schmid and R.W. Spekkens, Contextual Advantage for State Discrimination, Phys. Rev. X \textbf{8}, 011015 (2018).
		
		\bibitem{PBR} M. F. Pusey, B. Jonathan, and R. Terry, On the reality of the quantum state, Nat. Phys. \textbf{8}, 475 (2012).
		
		\bibitem{QKD} C. H. Bennett, Quantum cryptography using any two nonorthogonal states, Phys. Rev. Lett. \textbf{68}, 3121 (1992).	
		
		\bibitem{QIP} J. Bae and L. C. Kwek, Quantum state discrimination and its applications, J. Phys. A \textbf{48}, 083001 (2015).
		
		\bibitem{NJP1} C. H. Bennett and G. Brassard, Quantum cryptography, Proc. IEEE Int. Conf. on Computers, Systems and Signal Processing, Bangalore, India. 175 (1984).
		
		\bibitem{NJP2} W. K. Wootters and W. H. Zurek, A single quantum cannot be cloned, Nature \textbf{299}, 802 (1982).
		
		\bibitem{NJP6}  N. Gisin, Quantum cloning without signaling, Phys. Lett. A \textbf{242}, 1 (1998).
		
		\bibitem{NJP18} N. Brunner, M. Navascués, and T. V{\'e}rtesi, Dimension witnesses and quantum state discrimination, Phys. Rev. Lett. \textbf{110}, 150501 (2013).
		
		\bibitem{NJP19} R. Koenig, R. Renner, and C. Schaffner, The operational meaning of min-and max-entropy, IEEE Trans. Inf. Theory \textbf{55}, 4337 (2009).
		
		\bibitem{Helstrom} C. W. Helstrom, Quantum Detection and Estimation Theory, J. Stat. Phys. \textbf{1}, 231 (1969).
		
		\bibitem{CHSH} J. F. Clauser, M. A. Horne, A. Shimony, and R. A. Holt, Proposed Experiment to Test Local Hidden-Variable Theories, Phys. Rev. Lett. {\bf 23}, 880 (1969).
		
		\bibitem{Olmschenk} S. Olmschenk, K. C. Younge, D. L. Moehring, D. N. Matsukevich, P. Maunz, and C. Monroe, Manipulation and detection of a trapped Yb hyperfine qubit,  Phys. Rev. A \textbf{76}, 052314 (2007).
		
		\bibitem{Detection} C. Shen and L. M. Duan,  Correcting detection errors in quantum state engineering through data processing,  \href{https://iopscience.iop.org/article/10.1088/1367-2630/14/5/053053/meta}{New J. Phys. \textbf{14}, 053053 (2012).}  
		
		\bibitem{supp} See Supplemental Materials for the data and error analysis.
		
		\bibitem{NC} M. D. Mazurek, M. F. Pusey, R. Kunjwal, K. J. Resch, and R. W. Spekkens, An experimental test of noncontextuality without unphysical idealizations, Nat. Commun. \textbf{7}, 11780 (2016).
		
		\bibitem{Mikeprxq} M. D. Mazurek, M. F. Pusey, K. J. Resch, and R. W. Spekkens, Experimentally bounding deviations from quantum theory in the landscape of generalized probabilistic theories, PRX Quantum \textbf{2}, 020302 (2021).
		
		\bibitem{RB} E. Knill, D. Leibfried, R. Reichle, J. Britton, R. B. Blakestad, J. D. Jost, C. Langer, R. Ozeri, S. Seidelin, and D. J. Wineland, Randomized benchmarking of quantum gates, Phys. Rev. A \textbf{77},  012307 (2008).		
		
	\end{thebibliography}
\end{document}